# Tunneling Anisotropic Magnetoresistance in Ferroelectric Tunnel Junctions

Artem Alexandrov,[1] M. Ye. Zhuravlev,[1,2*] and Evgeny Y. Tsymbal[1,3†]

[1] *Moscow Institute of Physics and Technology, Dolgoprudny, Moscow Region 141700, Russia*
[2] *St. Petersburg State University, St. Petersburg 190000, Russia*
[3] *Department of Physics and Astronomy, University of Nebraska, Lincoln, Nebraska 68588, USA*

Using a simple quantum-mechanical model, we explore a tunneling anisotropic magnetoresistance (TAMR) effect in ferroelectric tunnel junctions (FTJs) with a ferromagnetic electrode and a ferroelectric barrier layer, which spontaneous polarization gives rise to the Rashba and Dresselhaus spin-orbit coupling (SOC). For realistic parameters of the model, we predict sizable TAMR measurable experimentally. For asymmetric FTJs, which electrodes have different work functions, the built-in electric field affects the SOC parameters and leads to TAMR dependent on ferroelectric polarization direction. The SOC change with polarization switching affects tunneling conductance, revealing a new mechanism of tunneling electroresistance (TER). These results demonstrate new functionalities of FTJs which can be explored experimentally and used in electronic devices.

## I. INTRODUCTION

Ferroelectric tunnel junctions (FTJ) have aroused considerable interest due to the interesting physics and potential applications as nanoscale resistive switching devices.[1-3] A FTJ consists of two metal electrodes separated by a nm-thick ferroelectric (FE) barrier which allows electron tunneling through it. The key property is the tunneling electroresistance (TER) effect that is a change in resistance of a FTJ with reversal of FE polarization. Following the theoretical predictions,[4,5] there have been a number of experimental demonstrations of the TER effect in trilayer junctions.[6-10] It was shown that the sizable TER effect can be achieved by using dissimilar electrodes[11-13], interface engineering[14-16], applied bias[17,18], or defect control.[19] Contrary to FE capacitors where leakage currents are detrimental to the device performance, the conductance of a FTJ is the functional characteristic of the device.[20] This makes FTJs promising for non-volatile memory applications.[21,22]

The functionality of FTJs can be extended by using ferromagnetic electrodes to create a multiferroic tunnel junction (MFTJ).[23] A MFTJ combines properties of a FTJ and a magnetic tunnel junction (MTJ), which exhibits a tunneling magnetoresistance (TMR) effect, that is the dependence of resistance on the relative magnetization directions in the two ferromagnetic electrodes.[24] Due to the coexistence of the TER and TMR effects, a MFTJ constitutes a four-state resistance device where the resistance is controlled both by the FE polarization direction of the barrier and the magnetization alignment of the ferromagnetic electrodes. MFTJs are interesting from the point of view of their multifunctional properties, as has been demonstrated in a number of experimental studies.[25-28]

In parallel with these developments, it has been found that resistance of MTJs can also depend on the magnetization orientation with respect to the crystallographic axes.[29-34] This phenomenon is known as tunneling anisotropic magneto-resistance (TAMR). TAMR is the manifestation of one of the oldest known effects that couple magnetism and electronic transport, i.e. anisotropic magnetoresistance (AMR),[35] in the tunneling regime. Both TAMR and AMR are driven by spin-orbit coupling (SOC), entangling the spin and orbital degrees of freedom. Contrary to the conventional TMR effect, TAMR may occur in MTJs with only one ferromagnetic electrode. This functionality opens new possibilities for spintronic devices.

Exploring a TAMR effect in a FTJ, which transport properties are strongly dependent on FE polarization, is interesting due to the interplay between ferroelectricity, magnetism, spin-dependent transport, and SOC. Recent studies have demonstrated the strong SOC effects in a number of bulk FE materials. Very large SOC ($\sim 10^2$-$10^3$ meV) has been predicted, resulting from a polarization-induced potential gradient.[36-43] In addition to the sizable SOC favorable for the experimental demonstration of the TAMR effect, these materials have the advantage of the reversible FE polarization which can be switched by an applied electric field. Since FE materials are non-centrosymmetric, the spin-momentum coupling linear in wave vector $k$ is allowed by symmetry, giving rise to the linear Rashba and Dresselhaus SOC in bulk of these compounds.[44] As a result, reversal of FE polarization changes the SOC parameters and thus can affect the TAMR, similar to the recently predicted effect of FE polarization of the tunneling anomalous Hall effect (TAHE).[45]

In this work, we explore the appearance of the TAMR effect in FTJs with a single ferromagnetic electrode and a FE barrier layer which spontaneous polarization gives rise to the Rashba and Dresselhaus SOC. For realistic SOC parameters, we predict the appearance of a sizable TAMR effect measurable experimentally. For FTJs with the electrodes of different work functions, TAMR depends on FE polarization direction due its effect on SOC parameters. The latter also affects tunneling resistance, which reveals a new mechanism of TER.

## II. THEORETICAL APPROACH

We consider a FTJ, which consists of a semi-infinite left (L) ferromagnetic (FM) electrode ($z < 0$) and a right (R) nonmagnetic (NM) electrode ($z > a$) separated by a ferroelectric (FE) barrier layer of thickness $a$ (Fig. 1 (a)). Magnetization **M** of the ferromagnet is assumed to lie in the plane of the layer at angle $\phi$ with respect to the $x$ axis.

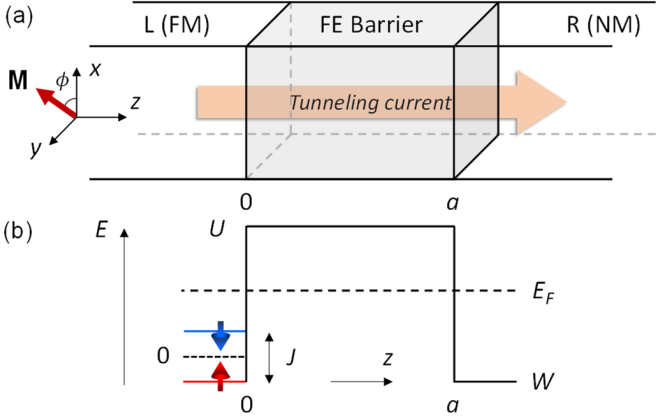

FIG. 1. (a) Schematic structure of a FTJ, which consists of semi-infinite left (L) ferromagnetic (FM) and right (R) nonmagnetic (NM) electrodes separated by a ferroelectric (FE) barrier of thickness $a$. Magnetization **M** lies in the $x$-$y$ plane at angle $\phi$ with respect to the x axis. (b) Potential profile across the junction. $E_F$ is the Fermi energy, $U$ is the barrier height, and $J_{ex}$ is the exchange splitting.

The corresponding Hamiltonian in each region is given by

$$\begin{cases} H_L = -\dfrac{\hbar^2}{2m}\nabla^2 - \dfrac{J}{2}\left(\sigma_x\cos\phi + \sigma_y\sin\phi\right), & z<0; \\ H_B = -\dfrac{\hbar^2}{2m}\nabla^2 + U + H_{SOC}, & 0<z<a; \\ H_R = -\dfrac{\hbar^2}{2m}\nabla^2 + W, & a>z. \end{cases} \quad (1)$$

Here $J$ is the exchange splitting in the FM electrode, $\sigma_x$ and $\sigma_y$ are the Pauli matrices, $m$ is the electron effective mass, $U$ is the barrier height, and $W$ is the potential in the NM electrode, as shown schematically in Fig. 1 (b). SOC in Eq. (1) is given by

$$H_{SOC} = \lambda_R(k_x\sigma_y - k_y\sigma_x) + \lambda_D(k_x\sigma_y + k_y\sigma_x), \quad (2)$$

where $\lambda_R$ and $\lambda_D$ are respectively the Rashba and Dresselhaus SOC constants. This form of SOC is typical for FE materials of the $C_{2v}$ point group where the polarization is pointing along the $z$-direction. The SOC Hamiltonian can be rewritten in terms of the effective **k**-dependent spin-orbit field $\mathbf{\Omega}(\mathbf{k})$ coupled to the spin as follows

$$H_{SOC} = \mathbf{\Omega}(\mathbf{k})\cdot\boldsymbol{\sigma}. \quad (3)$$

Comparing to Eq. (2), it is seen from that $\mathbf{\Omega}(\mathbf{k}) = (\beta k_y, \alpha k_x, 0)$, where $\alpha = \lambda_R + \lambda_D$ and $\beta = \lambda_D - \lambda_R$.

To calculate the conductance, we use the standard approach based on transmission of the incoming electron wave across the junction. The propagating state of energy $E$ incoming from the left FM electrode and normalized to the unit current density is given by

$$\psi_L^\sigma = \sqrt{\dfrac{m}{\hbar k_z^\sigma}}e^{i(k_x x + ik_y y)}e^{ik_z^\sigma z}\chi_\phi^\sigma, \quad (4)$$

where $\sigma = \uparrow,\downarrow$ is the spin index, $k_z^{\uparrow,\downarrow} = \sqrt{2m(E\pm J/2)/\hbar^2 - k_\parallel^2}$ is the $z$-component of the wave vector in the FM electrode, $\mathbf{k}_\parallel = (k_x,k_y)$ is the transverse wave vector, which is conserved in the process of tunneling, and $\chi_\phi^{\uparrow,\downarrow} = \dfrac{1}{\sqrt{2}}\begin{pmatrix} e^{-i\phi/2} \\ \pm e^{i\phi/2} \end{pmatrix}$ is the spinor eigenfunction. The scattering state in the right electrode due to transmission of $\psi_L^\sigma$ across the junction is given by

$$\psi_{R\leftarrow L}^\sigma = \sqrt{\dfrac{m}{\hbar q_z}}\left(t_{RL}^{\sigma\sigma}e^{iq_z z}\chi_\phi^\sigma + t_{RL}^{\bar\sigma\sigma}e^{iq_z z}\chi_\phi^{\bar\sigma}\right), \quad (5)$$

where $\bar\sigma = -\sigma$, $q_z = \sqrt{2m(E-W)/\hbar^2 - k_\parallel^2}$, and $t_{RL}^{\bar\sigma\sigma}$ and $t_{RL}^{\sigma\sigma}$ are the transmission amplitudes with and without spin flip, respectively. The scattering state in the left electrode due to reflection of $\psi_L^\sigma$ from the barrier can be written as

$$\psi_{L\leftarrow L}^\sigma = \sqrt{\dfrac{m}{\hbar k_z^\sigma}}\left(e^{iq_z z} + r_{LL}^{\sigma\sigma}e^{-ik_z^\sigma z}\right)\chi_\phi^\sigma + \sqrt{\dfrac{m}{\hbar k_z^{\bar\sigma}}}r_{LL}^{\bar\sigma\sigma}e^{-ik_z^{\bar\sigma} z}\chi_\phi^{\bar\sigma}, \quad (6)$$

respectively, where $r_{LL}^{\sigma\bar\sigma}$ and $r_{LL}^{\sigma\sigma}$ are the reflection amplitudes with and without spin flip, respectively.

The scattering state in the barrier is given by

$$\psi_B = e^{i(k_x x + k_y y)}\begin{pmatrix} a_1^+ e^{Q_+ z} + a_2^+ e^{-Q_+ z} + a_1^- e^{Q_- z} + a_2^- e^{-Q_- z} \\ \gamma(a_1^+ e^{Q_+ z} + a_2^+ e^{-Q_+ z} - a_1^- e^{Q_- z} - a_2^- e^{-Q_- z}) \end{pmatrix}, \quad (7)$$

where $Q_\pm = \sqrt{2m(U-E\pm w)/\hbar^2 + k_\parallel^2}$, $\gamma = (i\alpha k_x + \beta k_y)/w$ and $w = \sqrt{\alpha^2 k_x^2 + \beta^2 k_y^2}$. The total conductance at zero temperature is given by

$$G = \int G(\mathbf{k}_\parallel)d\mathbf{k}_\parallel = \dfrac{e^2}{(2\pi)^3\hbar}\sum_\sigma\int\left(|t_{LR}^{\sigma\sigma}|^2 + |t_{LR}^{\bar\sigma\sigma}|^2\right)_{E=E_F}d\mathbf{k}_\parallel. \quad (8)$$

The respective transmission amplitudes are obtained by matching the wave functions given by Eqs. (4)-(7) at the FTJ interfaces.

## III. RESULTS AND DISCUSSION

Next, we perform numerical calculations of the conductance for different magnetization angles. In the calculations, we assume $a = 3$ nm, $E_F = 3$ eV, $U = 1$ eV, $W = -10$ eV, and $J = 3$ eV as representative parameters. We start from analyzing the $\mathbf{k}_\parallel$-resolved conductance $G(\mathbf{k}_\parallel)$ given by Eq. (8) to demonstrate how the specific form of SOC affects the conductance and its dependence on the magnetization angle.

First, we assume a purely Rashba-type SOC in Eq. (3), i.e. $\lambda_R = 1$ eV·Å and $\lambda_D = 0$. Figure 2(a) shows the calculated $\mathbf{k}_\parallel$-resolved conductance for different magnetization angles $\phi$. It is seen that at certain $\mathbf{k}_\parallel$ the conductance exhibits a maximum, which follows the changing magnetization direction. This behavior can be understood as follows. The Rashba SOC in the FE layer leads to the splitting of the free electron band, which determines the effective $\mathbf{k}_\parallel$-dependent barrier height for tunneling electrons:

$$U_{eff}^{\uparrow,\downarrow} = U - E_F + \frac{\hbar^2 k_\parallel^2}{2m} \pm \lambda_R k_\parallel . \quad (9)$$

The lowest barrier height occurs in a circular region around the origin at $k_\parallel = m\lambda/\hbar^2$, which is about 0.13 Å for $\lambda = 1$ eV·Å. The spin follows the effective SOC field $\mathbf{\Omega} = \lambda_R(-k_y, k_x, 0)$, pointing perpendicular to the wave vector $\mathbf{k}_\parallel$ and creating a circular spin texture schematically shown by arrows in Fig. 2 (a). For those $\mathbf{k}_\parallel$, where the spin points parallel to the magnetization (indicated by red arrows in Fig. 2 (a)), the transmission is largest. This is due to the majority-spin electrons, incoming from the FM layer, encountering the lowest barrier height. For other $\mathbf{k}_\parallel$, where the spin is misaligned to the magnetization, the transmission is less efficient due to the admixture of the minority-spin states. When the magnetization angle $\phi$ changes, the highest transmission region rotates with the magnetization due to the circular spin texture induced by the Rashba SOC. A qualitatively similar behavior occurs for a purely Dresselhaus-type SOC (not shown), when $\lambda_D = 1$ eV·Å and $\lambda_R = 0$. The difference is in the conductance maxima locations and their sense of rotation with the changing angle $\phi$.

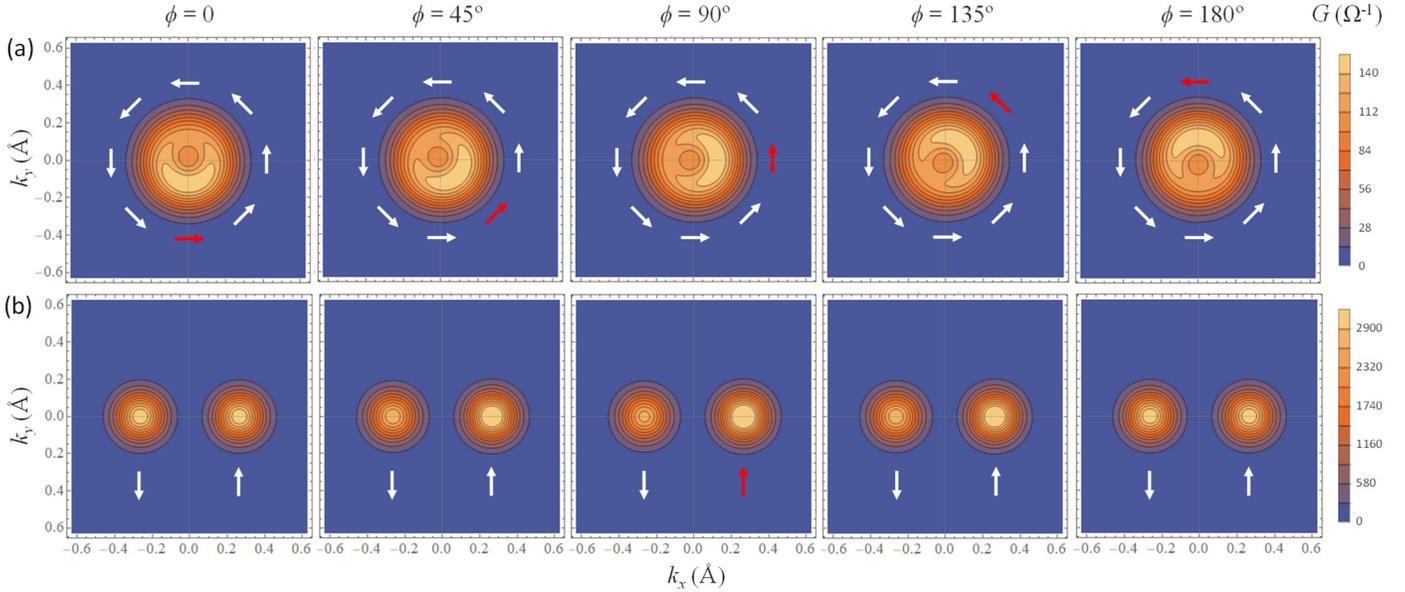

FIG. 2. $\mathbf{k}_\parallel$-resolved conductance $G(k_x, k_y)$ for different magnetization angles $\phi$ for $\lambda_R = 1$ eV·Å and $\lambda_D = 0$ (a) and $\lambda_R = \lambda_D = 1$ eV·Å (b). Arrows indicate spin-orbit field orientation. Red color specifies the spin-orbit field being parallel to magnetization.

In the opposite limit, when $\lambda_R$ and $\lambda_D$ are equal, e.g. $\lambda_D = \lambda_R = 1$ eV·Å, the spin-orbit field is unidirectional, i.e. $\mathbf{\Omega} = \alpha(0, k_x, 0)$, forming a momentum-independent spin configuration known as the persistent spin texture.[43] In this case, the effective tunneling barrier height is given by

$$U_{eff}^{\uparrow,\downarrow} = U - E_F + \frac{\hbar^2 k_\parallel^2}{2m} \pm (\lambda_R + \lambda_D) k_x , \quad (10)$$

and the lowest barrier height occurs at $k_y = 0$ and $k_x = \pm m(\lambda_R + \lambda_D)/\hbar^2$ which is about ±0.26 Å for $\lambda_R = \lambda_D = 1$

eV·Å. The conductance maxima occur at these $\mathbf{k}_\parallel$ points with the relative weight being dependent on the magnetization angle $\phi$ (Fig. 2 (b)). For $\phi = 0$, the effective SOC field is pointing along the *y*-axis and thus the magnetization is normal to the spin direction (shown by arrows in Fig. 2 (b)), resulting in the equal conductance at the two maxima. However, for $\phi = \pi/2$, the magnetization is parallel to the spin at $k_x = m(\lambda_R + \lambda_D)/\hbar^2$ (indicated by the red arrow in Fig. 2 (b)) and antiparallel to the spin at $k_x = -m(\lambda_R + \lambda_D)/\hbar^2$, leading to the higher conductance for the former.

the amplitude of TAMR as a function of the SOC constants and its angular dependence are well described by a simple formula

$$g(\phi) \propto \lambda_R \lambda_D \sin^2 \phi. \quad (11)$$

This formula follows from the expression for the conductance derived using second order perturbation theory for SOC.[46]

Figure 4 (b) shows the TAMR amplitude $\eta$ as a function of the exchange constant $J$ in the FM layer for $\lambda_R = \lambda_D = 1$ eV·Å. As expected, $\eta$ increases with $J$ from zero for to finite values due to the increasing disbalance between the number of majority- and minority-spin carriers in the FM electrode.

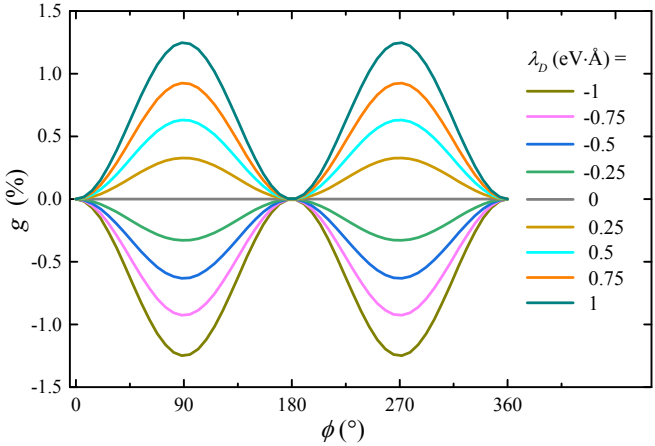

FIG. 3. Normalized conductance $g(\phi) = \dfrac{G(\phi) - G(0)}{G(0)}$ as a function of magnetization angle $\phi$ for $\lambda_R = 1$ eV·Å and different values of $\lambda_D$.

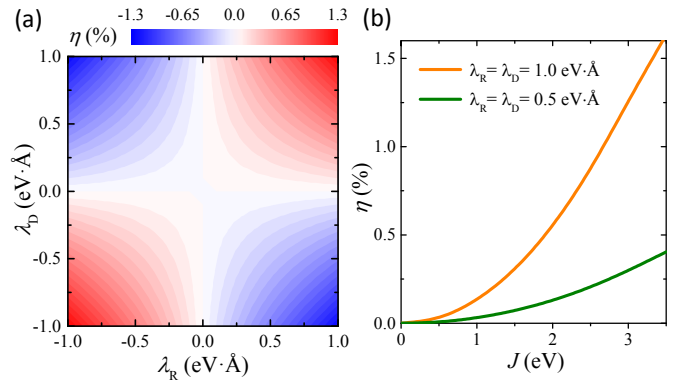

FIG. 4. Amplitude of TAMR $\eta = (G(\pi/2) - G(0))/G(0)$ as a function of $\lambda_R$ and $\lambda_D$ (a) and exchange constant $J$ for $\lambda_R = \lambda_D = 1$ eV·Å (orange line) and $\lambda_R = \lambda_D = 0.5$ eV·Å (green line) (b).

The observed changes in the $\mathbf{k}_\parallel$-conductance with magnetization angle are expected to produce a TAMR effect. For $\lambda_D = 0$ (or $\lambda_R = 0$), however, the rotational symmetry of the band structure and the spin texture (Fig. 2(a)) lead to the vanishing of TAMR, where the total conductance does not depend on angle $\phi$ (Fig. 3, grey line). Only when both $\lambda_R$ and $\lambda_D$ are non-zero, the TAMR effect is finite as revealed in Figure 3 by the oscillatory variation of the normalized conductance $g(\phi) = \dfrac{G(\phi) - G(0)}{G(0)}$ with the magnetization angle. This variation is well described by $g(\phi) \propto \sin^2 \phi$ typical for AMR is general. For a given value of $\lambda_R$ (e.g., $\lambda_R = 1$ eV·Å in Fig. 3), the amplitude of TAMR, which is defined as $\eta = g(\pi/2)$, increases with $\lambda_D$ in a linear fashion (Fig. 3).

Figure 4 (a) shows the calculated amplitude of TAMR as a function of $\lambda_R$ and $\lambda_D$. We find that this variation quantitatively follows the relationship $\eta \propto \lambda_R \lambda_D$. Overall, both

As follows from the discussion above, the TAMR effect in FTJs is driven by the SOC in the FE barrier layer due to broken inversion symmetry, producing linear in wave vector $\mathbf{k}$ Rashba and Dresselhaus SOC effects. Switching FE polarization $\mathbf{P}$ in bulk ferroelectric leads to a full reversal of the spin texture is as was originally proposed for GeTe.[36] This is due to the fact that a change of $\mathbf{P}$ to $-\mathbf{P}$ is equivalent to the space inversion operation which changes the wave vector from $\mathbf{k}$ to $-\mathbf{k}$ but preserves the spin $\sigma$. Applying the time-reversal symmetry operation to this state with reversed polarization, we transform $-\mathbf{k}$ back $\mathbf{k}$ but flip the spin, changing it from $\sigma$ to $-\sigma$. Thus, the reversed-polarization state is identical to the original state with the same $\mathbf{k}$ but reversed $\sigma$. For a FTJ, this property has been predicted to be responsible for the change of sign of the TAHE.[45]

In case of TAMR, however, the situation is different. Due to the TAMR effect being a second order in SOC (the first-order term in the expansion is an odd function of $\mathbf{k}$ and thus vanishes after integration over $\mathbf{k}_\parallel$),[46] reversal of FE polarization leading to the reversal of spin at a given $\mathbf{k}_\parallel$ does not change the conductance and hence the TAMR. Nevertheless, in realistic

FTJs, the effect of polarization switching on TAMR is expected to occur due to the intrinsic asymmetry of FTJs. This asymmetry is known to appear as a result of different screening lengths and/or work functions of the electrodes or different interface terminations, leading to the TER effect.[3]

To illustrate the effect of asymmetry, we assume that a FTJ has FM and NM electrodes with different work functions, resulting in a built-in electric field. This electric field, pointing in a fixed direction, is expected to affect the SOC parameters when the FE polarization is reversed and thus to change TAMR. If $\Delta W$ is the difference in work functions of the two electrodes, the induced electric field in a FE barrier layer is given by $E_i = \Delta W / ea$, where $e$ is an elementary charge and $a$ is barrier thickness. This field induces polarization $P_i = E_i(\epsilon - \epsilon_0)$, where $\epsilon$ is the background dielectric permittivity of the ferroelectric. The induced polarization has a fixed orientation and hence adds to or subtracted from the spontaneous polarization $P$. The total polarization is given by $P + P_i$ or $-P + P_i$ depending on the spontaneous polarization being parallel or antiparallel to the built-in electric field, respectively.

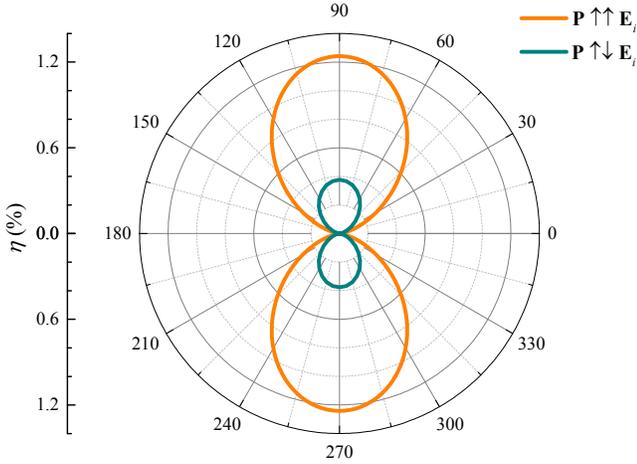

FIG. 5. Conductance of a FTJ as a function of magnetization angle $\phi$ for polarization **P** parallel ($\lambda_R = \lambda_D = 1$ eV·Å, orange line) and antiparallel ($\lambda_R = \lambda_D = -0.56$ eV·Å, cyan line) to the built-in electric field **E**$_i$.

For simplicity, we assume that the spin-orbit field and the SOC constants are proportional to the total polarization $P$.[47] If the spontaneous polarization $P$ of bulk ferroelectric gives rise to the SOC constants $\lambda_{R(D)}$, in a FTJ with the induced polarization $P_i$ due to the built-in electric field, these constants become $\lambda_{R(D)}^{\pm} = \pm \lambda_{R(D)}(P \pm P_i)/P$, where sign + (–) designates polarization pointing parallel (antiparallel) to the built-in electric field.

For an estimate, we consider representative parameters $\Delta W = 1$ eV, $a = 2$ nm, and $\epsilon = 30\epsilon_0$, which lead to the built-in field $E_i = 0.5$ V/nm and induced polarization $P_i \approx 14$ μC/cm$^2$. If the spontaneous polarization is $P = 50$ μC/cm$^2$, the total polarization is $P + P_i = 64$ μC/cm$^2$ or $-P + P_i = -36$ μC/cm$^2$ depending on its orientation. This nearly twofold change in the absolute value of the polarization is mirrored in the change of the SOC parameters. For example, if $\lambda_{R(D)}^{+} = 1$ eV·Å (i.e. $\lambda_{R(D)} \approx 0.78$ eV·Å) then $\lambda_{R(D)}^{-} \approx -0.56$ eV·Å. Thus, reversal of FE polarization leads to a nearly twofold change in the SOC constants, which according to Eq. (11) is expected to produce about fourfold change in TAMR. Figure 5 shows the calculated TAMR curves in polar coordinates, which illustrate this behavior.[48]

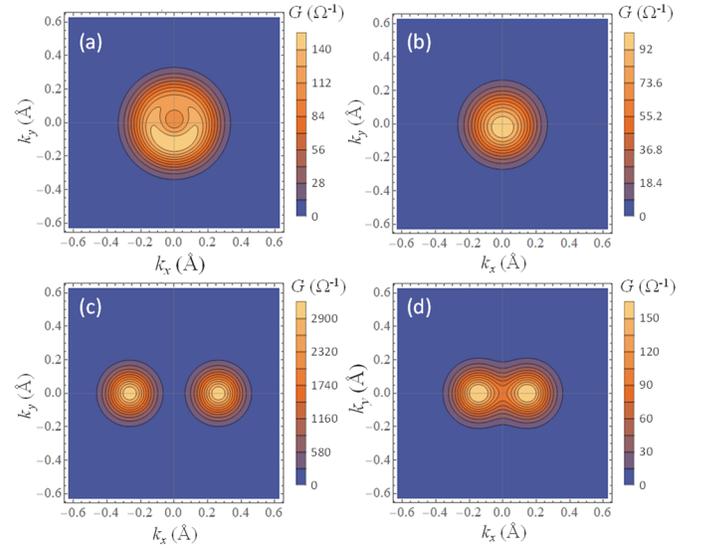

FIG. 6. $\mathbf{k}_\parallel$-resolved conductance for $\lambda_R = 1$ eV·Å and $\lambda_D = 0$ (a), $\lambda_R = -0.56$ eV·Å and $\lambda_D = 0$ (b), $\lambda_R = \lambda_D = 1$ eV·Å (c) and $\lambda_R = \lambda_D = -0.56$ eV·Å (d). The magnetization angle is fixed at $\phi = 0$.

An interesting observation which follows from our calculations is a sizable effect of the SOC parameters on conductance. Since the SOC parameters are affected by the orientation of FE polarization, we argue that this manifests a new mechanism of TER. Figures 6 (a, b) show the calculated $\mathbf{k}_\parallel$-resolved conductance for purely Rashba SOC ($\lambda_D = 0$), when it changes from $\lambda_R = 1$ eV·Å (Fig. 6 (a)) to $\lambda_R = -0.56$ eV·Å (Fig. 6 (b)) with reversal of FE polarization. We find that the total conductance of the FTJ changes from about 27.4 Ω$^{-1}$ cm$^{-2}$ to 9.9 Ω$^{-1}$ cm$^{-2}$. The effect stems entirely from changes in the electronic band structure of the FE barrier driven by SOC. As evident from comparison of Figures 6 (a) and 6 (b), the reduction in the Rashba SOC shrinks the $\mathbf{k}_\parallel$-space where the conductance is sizable, and also, according to Eq. (9), enhances the barrier height. The effect is even more pronounced for $\lambda_R = \lambda_D$ (Figs. 6 (c, d)). When the SOC parameters change from 1 eV·Å (Fig. 6 (c)) to –0.56 eV·Å (Fig. 6 (d)), the conductance minima become closer, and the effective barrier height, according to Eq. (10), is reduced (stronger than in the case of pure Rashba SOC) resulting in the conductance change from about 318 $\Omega^{-1}$cm$^{-2}$ to 19.1 $\Omega^{-1}$cm$^{-2}$. Such a large change in the conductance corresponds to the ON/OFF ratio of about 17, which is comparable to the TER values predicted earlier.[4]

There are a number of FE materials that can be utilized in FTJs as barrier layers to realize the predicted properties experimentally. Among them is orthorhombic HfO$_2$, which was predicted to have a large Dresselhaus SOC, $\lambda_D$ = 0.578 eV·Å (though a moderate Rashba SOC, $\lambda_R$ = 0.028 eV·Å).[41] Recently FTJs based on polycrystalline FE HfO$_2$ films have been realized.[49-51] The experimental challenge, however, is to grow monocrystalline FE thin films of HfO$_2$, as required for observing the TAMR effect. Recently a rhombohedral FE phase was demonstrated in epitaxially-strained Hf$_{0.5}$Zr$_{0.5}$O$_2$ thin films with nearly monocrystalline quality,[52] which is promising for the experimental realization of TAMR.

Very interesting materials for the use as tunnel barriers in FTJs to observe the TAMR effect are those which maintain a persistent spin texture.[43] Among them is BiInO$_3$ which belongs to the *Pna*2$_1$ orthorhombic phase (space group No. 33) and has band structure described by the effective SOC parameters $\lambda_R = \lambda_D$ = 0.955 eV·Å. Growth of epitaxial thin films of BiInO$_3$ has not yet been reported, but given the very large SOC in this material, interesting effects including TAHE and TAMR could observed in FTJs based on this material.

## IV. CONCLUSIONS

In conclusion, using a simple quantum-mechanical model, we have explored the TAMR effect in FTJs with a FM electrode. The effect is driven by SOC in the FE tunneling barrier, which has two contributions: the bulk Rashba SOC and the linear in wave vector **k** Dresselhaus SOC. By calculating the $\mathbf{k}_\parallel$-resolved conductance for different magnetization angles, we analyzed the effect of the SOC. For realistic parameters of the model, we found sizable TAMR values, which can be measured experimentally. We argued that a built-in electric field in asymmetric FTJs affects the SOC parameters and leads to a change in TAMR when FE polarization of the barrier layer is reversed. This prediction was elaborated, using an example of a FTJ with electrodes of different work functions. Finally, we found that a change in the SOC parameters with polarization switching leads to a sizable change in conductance, which manifests a new mechanism of TER. We hope that our theoretical predictions will stimulate experimental studies of TAMR in FTJs.

## ACKNOWLEDGMENTS

The authors thank Drs. Andrei Zenkevich and Lingling Tao for helpful discussions. This work was financially supported by the Moscow Institute of Physics and Technology (MIPT) and by the Russian Science Foundation (Grant No. 18-12-00434).

* myezhur@gmail.com  
† tsymbal@unl.edu

[1] E. Y. Tsymbal and H. Kohlstedt, Tunneling across a ferroelectric. *Science* **313**, 181 (2006).

[2] V. Garcia and M. Bibes, Ferroelectric tunnel junctions for information storage and processing. *Nat. Comm.* **5**, 4289 (2014).

[3] J. P. Velev, J. D. Burton, M. Y. Zhuravlev, and E. Y. Tsymbal, Predictive modelling of ferroelectric tunnel junctions. *npj Comp. Mater.* **2**, 16009 (2016).

[4] M. Y. Zhuravlev, R. F. Sabirianov, S. S. Jaswal, and E. Y. Tsymbal, Giant electroresistance in ferroelectric tunnel junctions. *Phys. Rev. Lett.* **94**, 246802 (2005).

[5] H. Kohlstedt, N. A. Pertsev, J. R. Contreras, and R. Waser, Theoretical current-voltage characteristics of ferroelectric tunnel junctions. *Phys. Rev. B* **72**, 125341 (2005).

[6] A. Chanthbouala, A. Crassous, V. Garcia, K. Bouzehouane, S. Fusil, X. Moya, J. Allibe, B. Dlubak, J. Grollier, S. Xavier, C. Deranlot, A. Moshar, R. Proksch, N. D. Mathur, M. Bibes, and A. Barthelemy, Solid-state memories based on ferroelectric tunnel junctions. *Nat. Nanotech.* **7**, 101 (2012).

[7] Z. Wen, C. Li, D. Wu, A. Li, and N. B. Ming, Ferroelectric-field-effect-enhanced electroresistance in metal/ferroelectric/semiconductor tunnel junctions. *Nat. Mater.* **12**, 617 (2013).

[8] H. Lu, A. Lipatov, S. Ryu, D. J. Kim, H. Lee, M. Y. Zhuravlev, C. B. Eom, E.Y. Tsymbal, A. Sinitskii, and A. Gruverman, Ferroelectric tunnel junctions with graphene electrodes. *Nat. Commun.* **5**, 5518 (2014).

[9] S. Boyn, A. M. Douglas, C. Blouzon, P. Turner, A. Barthelemy, M. Bibes, S. Fusil, J. M. Gregg, and V. Garcia, Tunnel electroresistance in BiFeO$_3$ junctions: size does matter. *Appl. Phys. Lett.* **109**, 232902 (2016).

[10] Z. Xi, J. Ruan, C. Li, C. Zheng, Z. Wen, J. Dai, A. Li, and D. Wu, Giant tunnelling electroresistance in metal/ferroelectric/semiconductor tunnel junctions by engineering the Schottky barrier. *Nat. Commun.* **8**, 15217 (2017).

[11] A. Zenkevich, M. Minnekaev, Yu. Matveyev, Yu. Lebedinskii, K. Bulakh, A. Chouprik, A. Baturin, K. Maksimova, S. Thiess, and W. Drube, Electronic band alignment and electron transport in Cr/BaTiO$_3$/Pt ferroelectric tunnel junctions. *Appl. Phys. Lett.* **102**, 062907 (2013).

[12] R. Soni, A. Petraru, P. Meuffels, O. Vavra, M. Ziegler, S. K. Kim, D. S. Jeong, N. A. Pertsev, and H. Kohlstedt, Giant electrode effect on tunnelling electroresistance in ferroelectric tunnel junctions. *Nat. Commun.* **5**, 5414 (2014).

[13] L. L. Tao and J. Wang, Ferroelectricity and tunneling electroresistance effect in asymmetric ferroelectric tunnel junctions. *J. Appl. Phys.* **119**, 224104 (2016).

[14] M. Y. Zhuravlev, Y. Wang, S. Maekawa, and E. Y. Tsymbal, Tunneling electroresistance in ferroelectric tunnel junctions with a composite barrier. *Appl. Phys. Lett.* **95**, 052902 (2009).

[15] A. Tsurumaki-Fukuchi, H. Yamada, and A. Sawa, Resistive switching artificially induced in a dielectric/ferroelectric composite diode. *Appl. Phys. Lett.* **103**, 152903 (2013).

[16] V. S. Borisov, S. Ostanin, S. Achilles, J. Henk, and I. Mertig, Spin-dependent transport in a multiferroic tunnel junction: Theory for Co/PbTiO$_3$/Co. *Phys. Rev. B* **92**, 075137 (2015).

[17] D. I. Bilc, F. D. Novaes, J. Íñiguez, P. Ordejón, and P. Ghosez, Electroresistance effect in ferroelectric tunnel junctions with symmetric electrodes. *ACS Nano* **6**, 1473 (2012).

[18] A. Useinov, A. Kalitsov, J. Velev, and N. Kioussis, Bias-dependence of the tunneling electroresistance and magnetoresistance in multiferroic tunnel junctions. *Appl. Phys. Lett.* **105**, 102403 (2014).

[19] K. Klyukin, L. L. Tao, E. Y. Tsymbal, and V. Alexandrov, Defect-assisted tunneling electroresistance in ferroelectric tunnel junctions. *Phys. Rev. Lett.* **121**, 056601 (2018).

[20] E. Y. Tsymbal and A. Gruverman, Ferroelectric tunnel junctions: Beyond the barrier *Nat. Mater.* **12**, 602 (2013).

[21] S. Boyn, S. Girod, V. Garcia, S. Fusil, S. Xavier, C. Deranlot, H. Yamada, C. Carrétéro, E. Jacquet, M. Bibes, A. Barthélémy, and J. Grollier, High-performance ferroelectric memory based on fully patterned tunnel junctions. *Appl. Phys. Lett.* **104**, 052909 (2014).

[22] M. Abuwasib, H. Lu, T. Li, P. Buragohain, H. Lee, C.-B. Eom, A. Gruverman, and U. Singisetti, Scaling of electroresistance effect in fully integrated ferroelectric tunnel junctions. *Appl. Phys. Lett.* **108**, 152904 (2016).

[23] M. Y. Zhuravlev, S. S. Jaswal, E. Y. Tsymbal, and R. F. Sabirianov, Ferroelectric switch for spin injection. *Appl. Phys. Lett.* **87**, 222114 (2005).

[24] M. Y. Zhuravlev, S. Maekawa, and E. Y. Tsymbal, Effect of spin-dependent screening on tunneling electroresistance and tunneling magnetoresistance in multiferroic tunnel junctions. *Phys. Rev. B* **81**, 104419 (2010).

[25] V. Garcia, M. Bibes, L. Bocher, S. Valencia, F. Kronast, A. Crassous, X. Moya, S. Enouz-Vedrenne, A. Gloter, D. Imhoff, C. Deranlot, N. D. Mathur, S. Fusi, K. Bouzehouane, and A. Barthélémy, Ferroelectric control of spin polarization. *Science* **327**, 1106 (2010).

[26] M. Hambe, A. Petraru, N. A. Pertsev, P. Munroe, V. Nagarajan, and H. Kohlstedt, Crossing an interface: Ferroelectric control of tunnel currents in magnetic complex oxide heterostructures. *Adv. Funct. Mater.* **20**, 2436 (2010).

[27] D. Pantel, S. Goetze, D. Hesse, and M. Alexe, Reversible electrical switching of spin polarization in multiferroic tunnel junctions. *Nat. Mater.* **11**, 289 (2012).

[28] Y. W. Yin, J. D. Burton, Y.-M. Kim, A. Y. Borisevich, S. J. Pennycook, S. M. Yang, T. W. Noh, A. Gruverman, X. G. Li, E. Y. Tsymbal, and Q. Li, Enhanced tunnelling electroresistance effect due to a ferroelectrically induced phase transition at a magnetic complex oxide interface. *Nat. Mater.* **12**, 397 (2013).

[29] M. Tanaka and Y. Higo, Large tunneling magnetoresistance in GaMnAs/AlAs/GaMnAs ferromagnetic semiconductor tunnel junctions. *Phys. Rev. Lett.* **87**, 026602 (2001).

[30] C. Gould, C. Rüster, T. Jungwirth, E. Girgis, G. M. Schott, R. Giraud, K. Brunner, G. Schmidt, and L. W. Molenkamp, Tunneling anisotropic magnetoresistance: A spin-valve-like tunnel magnetoresistance using a single magnetic layer. *Phys. Rev. Lett.* **93**, 117203 (2004).

[31] H. Saito, S. Yuasa, and K. Ando, Origin of the tunnel anisotropic magnetoresistance in Ga$_{1-x}$Mn$_x$As/ZnSe/Ga$_{1-x}$Mn$_x$As magnetic tunnel junctions of II-VI/III-V heterostructures. *Phys. Rev. Lett.* **95**, 086604 (2005).

[32] J. Moser, A. Matos-Abiague, D. Schuh, W. Wegscheider, J. Fabian, and D. Weiss, Tunneling anisotropic magnetoresistance and spin-orbit coupling in Fe/GaAs/Au tunnel junctions. *Phys. Rev. Lett.* **99**, 056601 (2007).

[33] A. N. Chantis, K. D. Belashchenko, E. Y. Tsymbal, and M. van Schilfgaarde, Tunneling anisotropic magnetoresistance driven by resonant surface states: First-principles calculations on an Fe (001) surface. *Phys. Rev. Lett.* **98**, 046601 (2007).

[34] A. Matos-Abiague and J. Fabian, Anisotropic tunneling magnetoresistance and tunneling anisotropic magnetoresistance: Spin-orbit coupling in magnetic tunnel junctions. *Phys. Rev. B* **79** (2009).

[35] W. Thomson, On the electro-dynamic qualities of metals: Effects of magnetization on the electric conductivity of nickel and of iron. *Proc. Roy. Soc. London* **8**, 546 (1857).

[36] D. Di Sante, P. Barone, R. Bertacco, and S. Picozzi, Electric control of the giant Rashba effect in bulk GeTe. *Adv. Mater.* **25**, 509 (2013).

[37] M. Kim, J. Im, A. J. Freeman, J. Ihm, and H. Jin, Switchable $S = 1/2$ and $J = 1/2$ Rashba bands in ferroelectric halide perovskites. *Proc. Natl. Acad. Sci. U.S.A.* **111**, 6900 (2014).

[38] A. Stroppa, P. Di Sante, P. Barone, M. Bokdam, G. Kresse, C. Franchini, M. H. Whangbo, and S. Picozzi, Tunable ferroelectric polarization and its interplay with spin–orbit coupling in tin iodide perovskites. *Nat. Commun.* **5**, 5900 (2014).

[39] L. L. Tao and J. Wang, Strain-tunable ferroelectricity and its control of Rashba effect in KTaO$_3$. *J. Appl. Phys.* **120**, 234101 (2016).

[40] L. G. D. da Silveira, P. Barone, and S. Picozzi, Rashba-Dresselhaus spin-splitting in the bulk ferroelectric oxide BiAlO$_3$. *Phys. Rev. B* **93**, 245159 (2016).

[41] L. L. Tao, T. R. Paudel, A. A. Kovalev, and E. Y. Tsymbal, Reversible spin texture in ferroelectric HfO$_2$. *Phys. Rev. B* **95**, 245141 (2017).

[42] J. He, D. Di Sante, R. Li, X. Q. Chen, J. M. Rondinelli, and C. Franchini, Tunable metal-insulator transition, Rashba effect and Weyl Fermions in a relativistic charge-ordered ferroelectric oxide. *Nat. Commun.* **9**, 492 (2018).

[43] L. L. Tao and E. Y. Tsymbal, Persistent spin texture enforced by symmetry. *Nat. Commun.* **9**, 2763 (2018).

[44] R. Winkler, *Spin-Orbit Coupling Effects in Two-Dimensional Electron and Hole Systems*, Springer Tracts in Modern Physics (Springer, Berlin, 2003).

[45] M. Y. Zhuravlev, A. Alexandrov, L. L. Tao, and E. Y. Tsymbal, Tunneling anomalous Hall effect in a ferroelectric tunnel junction. *Appl. Phys. Lett.* **113**, 172405 (2018).

[46] A. Matos-Abiague, M. Gmitra, and J. Fabian, Angular dependence of the tunneling anisotropic magnetoresistance in magnetic tunnel junctions. *Phys. Rev. B* **80**, 045312 (2009).

[47] H. Lee, J. Im, and H. Jin, Harnessing the giant out-of-plane Rashba effect and the nanoscale persistent spin helix via ferroelectricity in SnTe thin films. *arXiv* 1712.06112v3 (2017).

[48] In this calculation, we ignored for simplicity variation of the potential barrier due to the electric field resulting from different work functions of the electrodes.

[49] F. Ambriz-Vargas, G. Kolhatkar, M. Broyer, A. Hadj-Youssef, R. Nouar, A. Sarkissian, R. Thomas, C. Gomez-Yáñez, M. A. Gauthier, and A. Ruediger, A complementary metal oxide semiconductor process-compatible ferroelectric tunnel junction. *ACS Appl. Mater. Interfaces* **9**, 13262 (2017).

[50] A. Chouprik, A. Chernikova, A. Markeev, V. Mikheev, D. Negrov, M. Spiridonov, S. Zarubin, and A. Zenkevich, Electron transport across ultrathin ferroelectric $Hf_{0.5}Zr_{0.5}O_2$ films on Si. *Microelectron. Eng.* **178**, 250 (2017).

[51] Y. Goh and S. Jeon, Enhanced tunneling electroresistance effects in HfZrO-based ferroelectric tunnel junctions by high-pressure nitrogen annealing. *Appl. Phys. Lett.* **113**, 052905 (2018).

[52] Y. Wei, P. Nukala, M. Salverda, S. Matzen, H. J. Zhao, J. Momand, A. S. Everhardt, G. Agnus, G. R. Blake, P. Lecoeur, B. J. Kooi, J. Íñiguez, B. Dkhil, and B. Noheda, A rhombohedral ferroelectric phase in epitaxially strained $Hf_{0.5}Zr_{0.5}O_2$ thin films. *Nat. Mater.* **17**, 1095 (2018).